\documentclass[conference]{IEEEtran}
\usepackage{hyperref}
\IEEEoverridecommandlockouts
\usepackage{cite}
\usepackage{amsmath,amssymb,amsfonts}
\usepackage{algorithmic}
\usepackage{graphicx}
\usepackage{textcomp}
\usepackage{xcolor}
\usepackage{comment}
\usepackage[nonumberlist]{glossaries}
\usepackage{multirow}
\usepackage[nonumberlist]{glossaries}
\usepackage{threeparttable}
\usepackage{graphicx} 
\usepackage{tikz}
\usepackage{pgfplots}
\pgfplotsset{compat=1.18}
\usepackage{subcaption}

\def\BibTeX{{\rm B\kern-.05em{\sc i\kern-.025em b}\kern-.08em
    T\kern-.1667em\lower.7ex\hbox{E}\kern-.125emX}}

\newacronym{DL}{DL}{Deep Learning}
\newacronym{CNN}{CNN}{Convolutional Neural Network}
\newacronym{BLER}{BLER}{Block Error Rate}
\newacronym{PTQ}{PTQ}{Post-Training Quantization}

\begin{document}

\title{Efficient Deep Neural Receiver with Post-Training Quantization}
\author{\IEEEauthorblockN{
SaiKrishna Saketh Yellapragada}
\IEEEauthorblockA{\textit{{Aalto University}}  \\
Espoo, Finland \\
saikrishna.yellapragada@aalto.fi}
\and
\IEEEauthorblockN{
Esa Ollila}
\IEEEauthorblockA{\textit{{Aalto University}}  \\
Espoo, Finland \\
esa.ollila@aalto.fi}
\and
\IEEEauthorblockN{
Mário Costa}
\IEEEauthorblockA{\textit{Nokia Technologies Oy.}  \\
Espoo, Finland \\
mario.costa@nokia.com}
}

\maketitle
\begin{abstract} 
Deep learning has recently garnered significant interest in wireless communications due to its superior performance compared to traditional model-based algorithms. Deep convolutional neural networks (CNNs) have demonstrated notable improvements in block error rate (BLER) under various channel models and mobility scenarios. However, the high computational complexity and resource demands of deep CNNs pose challenges for deployment in resource-constrained edge systems. The 3rd Generation Partnership Project (3GPP) Release 20 highlights the pivotal role of artificial intelligence (AI) integration in enabling advanced radio-access networks for 6G systems. The hard real-time processing demands of 5G and 6G require efficient techniques such as post-training quantization (PTQ), quantization-aware training (QAT), pruning, and hybrid approaches to meet latency requirements. In this paper, we focus on PTQ to reduce model complexity by lowering the bit-width of weights, thereby enhancing computational efficiency. Our analysis employs symmetric uniform quantization, applying both per-tensor and per-channel PTQ to a neural receiver achieving performance comparable to full-precision models. Specifically, 8-bit per-channel quantization maintains BLER performance with minimal degradation, while 4-bit quantization shows great promise but requires further optimization to achieve target BLER levels. These results highlight the potential of ultra-low bitwidth PTQ for efficient neural receiver deployment in 6G systems.

\end{abstract}

\begin{IEEEkeywords}
deep learning, convolutional neural networks, wireless receiver, post-training quantization
\end{IEEEkeywords}

\section{Introduction} 
\label{sec:introduction}
Deep learning (DL) has significantly advanced numerous fields, demonstrating particular success in domains such as computer vision and language modeling. Recent research has shown that wireless communications systems benefit substantially from DL approaches when applied to the physical layer\cite{deeprx_1} \cite{e2e_faa}. These techniques have the potential of improved performance compared with model-based transceivers in terms of throughput and block error rate (BLER). The adoption of DL in wireless communications presents unique implementation challenges in terms of hardware efficiency and energy consumption at the edge devices. In addition, model training requires substantial time and hardware resources, significantly increasing the cost of DL receivers.
Previous work in efficient DL receivers by \cite{on_implem_nrx} examined a SISO-OFDM neural network-based receiver and demonstrated that uniform linear quantization-based QAT reduced floating-point operations (FLOPs) by 50\% while incurring only a 0.25 dB degradation in radio performance. In this paper, we design a lighter CNN-based deep learning receiver architecture proposed in \cite{deeprx_1}, and present a modified design aimed at reducing the memory size of the model, energy efficiency and improving latency during inference. The code developed in this paper is available on GitHub.\footnote{\label{git_repo}WIP: \url{https://github.com/saiksaketh/efficient_neural_rx}}

In this paper, we investigate the impact of post-training quantization (PTQ) on radio performance. PTQ reduces the precision of weights and activations in a pre-trained floating-point model (e.g., from \texttt{float32} to \texttt{float16/int8/int4} without requiring retraining). Through careful calibration, this technique preserves model accuracy with minimal performance degradation while delivering a more lightweight model that executes inference operations with greater speed on a hardware that supports the quantization operations. There are several PTQ variants, such as weight-only quantization or quantization of both activations and weights. Group quantization in PTQ achieves a balance between quantization accuracy and hardware efficiency\cite{lectures_songhan}. Techniques such as AdaRound may be used for improved PTQ by adaptive rounding of weights \cite{up_or_down}. Quantization-aware training (QAT) incorporates quantization effects during training and typically produces better performance than PTQ \cite{krishnamoorthi} \cite{jacob2017quantizationtrainingneuralnetworks}. In particular, QAT inserts simulated quantization operations into the neural network during the training process, allowing the model to learn optimal parameter values while accounting for the quantization effects that will be present during inference. This approach enables the network to adapt its weights to compensate for quantization errors, resulting in significantly better accuracy preservation at lower bit precision compared to post-training methods, particularly for aggressive quantization scenarios such as sub-8-bit representations. Pruning is a model optimization technique that removes redundant or less significant parameters (e.g., weights or entire filters in a CNN) to improve computational complexity and memory requirements while aiming to preserve model accuracy\cite{song_han_1}. In this paper, we focus on neural network receivers when subjected to PTQ, evaluating their radio performance, energy efficiency, and memory footprint. \par
The paper is structured as follows. Section \ref{sec:system_model} introduces the system model and the neural receiver architecture along with training and fine-tuning details. Section \ref{sec:ptq} discusses the principles of PTQ in detail, and variants in PTQ. Section \ref{sec:sim_results} shows the simulation results and BLER analysis for various user-equipment (UE) mobility conditions. Finally, conclusion and future work are given in Section \ref{sec:conclusions}.

\begin{table}[t]

\centering
\caption{Architecture Details of the Neural Receiver\label{table:arch_nrx}}
\begin{tabular}{| l | c | c | c |}
    \hline
    Layer & Channels & Kernel Size & Dilatation Rate \\ \hline
    Input Conv2D & 128 & (3,3) & (1,1) \\ \hline
    ResNet Blocks (1--8) & 128 & (3,3) & (1,1) \\ \hline
    Output Conv2D & 6 & (3,3) & (1,1) \\ \hline
\end{tabular}
\end{table}

\begin{table}[t]

\centering
\caption{Training Parameters and Randomization}
\label{table:nrx_training}
\begin{tabular}{|c|c|c|}
\hline
\textbf{Parameters} & \textbf{Training} & \textbf{Randomization} \\
\hline
Carrier Frequency & 3.5 GHz & \textit{None} \\
\hline
Channel Model & CDL-[A,B,C] & Uniform \\
\hline
RMS Delay Spread & 10 ns – 100 ns & Uniform \\
\hline
UE Velocity & 0 m/s – 50 m/s & Uniform \\
\hline
SNR & -2 dB – 15 dB & Uniform \\
\hline
Subcarrier Spacing & 30 kHz & \textit{None} \\
\hline
Modulation Scheme & 64-QAM & \textit{None} \\
\hline
No. of Tx Antennas & 1 & \textit{None} \\
\hline
No. of Rx Antennas & 2 & \textit{None} \\
\hline
Code Rate & 0.5 & \textit{None} \\
\hline
DMRS Configuration & 3\textsuperscript{rd} and 12\textsuperscript{th} Symbol & \textit{None} \\
\hline
Optimizer & Adam & \textit{None} \\
\hline
Batch Size & 128 & \textit{None} \\
\hline
\end{tabular}%
\label{tab:training_parameters}
\end{table}

\begin{table}[t]
\centering
\caption{Testing Parameters}\label{table:nrx_eval}
\begin{tabular}{|c|c|}
\hline
\textbf{Parameters} & \textbf{Tested on} \\
\hline
Carrier Frequency & 3.5 GHz \\
\hline
Channel Model & CDL-[D,E] \\
\hline
RMS Delay Spread & 0–50 ns \\
\hline
UE Velocity & 
\begin{tabular}[c]{@{}l@{}}
Low-speed: 0–5.1 m/s \\
Medium-speed: 10–20 m/s \\
High-speed: 25–40 m/s
\end{tabular} \\
\hline
SNR & 0 dB – 10 dB \\
\hline
\end{tabular}
\label{tab:test_parameters}
\end{table}

\section{System Model}
\label{sec:system_model}
The wireless communication system considered in this paper is based on Sionna \cite{sionna}. Sionna is a hardware-accelerated differentiable open-source library for research on communication systems. We consider a single-input and multiple-output (SIMO) system. The transmitter sends a low-density parity-check (LDPC) encoded bit sequence. The resulting code word is mapped into OFDM symbols and they are arranged on physical resource blocks (PRBs). Demodulation Reference Signal (DMRS) are inserted into designated sub-carriers and OFDM symbols to enable channel estimation at the receiver. Subsequently, data and pilots are converted into an OFDM waveform through inverse Fast Fourier transform (IFFT), and transmitted via the wireless channel. At the receiver, after additive measurement noise and FFT, the received signal can be expressed as (following the notation of ~\cite{deeprx_1}):

\begin{equation}
\label{eq1}
    \mathbf{y}_{ij} = \mathbf{h}_{ij} x_{ij} + \textbf{n}_{ij}.
\end{equation}  
Here, \( i \) and \( j \) denote the $i^{th}$ OFDM symbol and $j^{th}$ subcarrier indices, respectively. Moreover, \( \mathbf{y}_{ij} \in \mathbb{C}^{N_r \times 1} \) and \( x_{ij} \in \mathbb{C} \) denote the received and transmitted symbols, respectively, and \( \mathbf{h}_{ij} \in \mathbb{C}^{N_r \times 1} \) is the multi-channel corresponding to the $i^{th}$ OFDM symbol over the $j^{th}$ subcarrier. Finally, \( \textbf{n}_{ij} \in \mathbb{C}^{N_r \times 1} \) is the noise-plus-interference signal, and \( N_r \) is the number of receive antennas.

\subsection{Neural Receiver Architecture}
\label{subsec:nrx_arch}
The receiver architecture considered in this paper is a convolutional residual neural network (ResNet) designed to replace traditional signal processing blocks including channel estimation, layer demapping, equalization, and symbol demapping. The architecture of the neural receiver is summarized in Table \ref{table:arch_nrx}. It operates on frequency-domain received baseband samples and outputs log-likelihood ratios (LLRs) for subsequent LDPC decoding. The architecture stems from the design proposed in~\cite{deeprx_1, e2e_faa}. In particular, differences include hyperparameters and the number of residual blocks. Implementation details can be found in the accompanying code repository \ref{git_repo}.

\subsection{Training and Fine-Tuning of Neural Receiver}
The neural receiver architecture considered in this paper was designed and trained using Sionna and Tensorflow frameworks without pre-existing weights. The training of the neural receiver considered the parameters on Table \ref{tab:training_parameters}. In particular, the clustered delay line (CDL) channel model representing non-line-of-sight channels (CDL-A, CDL-B, and CDL-C) were employed for training. Testing was conducted on line-of-sight channels (CDL-D and CDL-E). During the first phase of training (up to $50,000$ iterations), a channel was selected among CDL-A, CDL-B, and CDL-C with a low-speed UE every $1,000$ iterations. This ensures robustness across diverse channel characteristics. 
During the fine-tuning phase, the model's performance was further refined. This stage spanned $30,000$ iterations. The velocity of UEs varied across low, medium, and high-speed every $10,000$ iterations. The SNR was chosen uniformly at random between $7$ and $12$ dB. This specific range was explored because it offered additional improvement in terms of BLER. The neural receiver was trained to maximize bit-metric decoding rate\cite{bmd}. 
The binary cross-entropy (BCE) loss is employed to measure the discrepancy between the predicted LLRs and the corresponding transmitted coded bits across the entire OFDM resource grid. The expected BCE is defined as

\begin{equation}
\small
\mathcal{L}_{\mathrm{BCE}}(B, \hat{L}) = - \mathbb{E}\left[B \log \sigma(\hat{L}) + (1 - B) \log (1 - \sigma(\hat{L}))\right],
\label{eq:batch_BCE_loss_training}
\end{equation}

where $B \in \{0,1\}$ denotes the ground-truth transmitted bits, $\hat{L}$ denotes the predicted LLRs and $\mathbb{E}[\cdot]$ expectation operator. The function $\sigma(\hat{L})$ represents the sigmoid activation, which maps LLRs to probabilities. 

Throughout the training process, L2 regularization with a coefficient of $1e^{-7}$ was implemented to mitigate overfitting and enhance training stability.

\section{Post-Training Quantization}
\label{sec:ptq}
This section summarizes the key mathematical aspects of quantization used in this paper. Integer quantization techniques are used to reduce the memory footprint of deep neural networks and improve inference latency. Compute-intensive tensor operations executed in \texttt{int8} precision can achieve up to a $16\times$ improvement in inference latency compared to their \texttt{float32} counterparts, depending on the employed hardware architecture \cite{wu2020integerquantizationdeeplearning}.

Quantization primarily involves two key operations: \textit{quantize}, where floating point values are assigned to a lower-bit integer representation (e.g. \texttt{int8}, \texttt{int4}), and \textit{dequantize}, where integer values are assigned back to approximate floating point values in the inference pipeline. The manner in which floating-point values are mapped to integer representations gives rise to two widely adopted strategies: asymmetric and symmetric quantization. Asymmetric quantization is defined by three quantization parameters: \textit{scale} ($s$), \textit{zero-point} ($z$), and \textit{bit-width} ($b$). The scale is a floating-point value that determines the step size of the quantizer, while the zero-point ensures that the real value zero is exactly representable in the integer domain.\par

In this paper we employ symmetric uniform quantization, where the zero-point is fixed at \( z = 0 \). This is a design choice based on the distribution of the data subject to quantization. The scale \( s \) is computed as:
\begin{equation}
s = \frac{\max\left(\left|x\right|\right) - \min\left(\left|x\right|\right)}{2^b - 1},
\label{eq:scale_computation}
\end{equation}
where \( \max(|x|) \) and \( \min(|x|) \) are the absolute maximum and minimum values of the tensor \( x \). 
The dequantization operation is defined as:
\begin{equation}
\hat{x} = s \cdot x_{\text{int}},
\label{eq:symmetric_quant}
\end{equation}
where $s$ is the scale and $x_{\text{int}}$ is the quantized integer value.

The quantization step for unsigned symmetric quantization is given by:
\begin{equation}
x_{\text{int}} = \text{clamp}\left( \left\lfloor \frac{x}{s} \right\rceil, 0, 2^b - 1 \right),
\label{eq:unsigned_quant}
\end{equation}
and for signed symmetric quantization:
\begin{equation}
x_{\text{int}} = \text{clamp}\left( \left\lfloor \frac{x}{s} \right\rceil, -2^{b-1}, 2^{b-1} - 1 \right),
\label{eq:signed_quant}
\end{equation}
where $\left\lfloor . \right\rceil$ is a round-to-nearest operator. Unsigned symmetric quantization is particularly suitable for one-tailed distributions such as ReLU activations, whereas signed symmetric quantization is more appropriate for zero-centered distributions \cite{qcom_white_paper}.

\subsection{Granularity of Quantization}
For a given tensor, \textit{per-tensor quantization} refers to the use of a single set of quantization parameters (scale, zero-point and bit-width) across the entire tensor. This technique assumes that all elements within the tensor share a similar dynamic range, making a single scaling factor and offset sufficient to map floating-point values to a lower bit-width representation (e.g., \texttt{int8}, \texttt{int4}). Per-tensor quantization is hardware-efficient since it allows a single set of parameters to be applied over a contiguous block of memory, simplifying memory access patterns.

In contrast, \textit{per-channel quantization} computes distinct quantization parameters for each output channel of a weight tensor, particularly in convolutional layers (e.g., Conv2D kernels). The channel-specific scale is defined as:

\begin{equation}
s_c = \frac{\max\left(\left|W[:,:,:,c]\right|\right) - \min(|W[:,:,:,c]|)}{2^b - 1}, \quad c = 1, \dots, C
\label{eq:per_channel_scale}
\end{equation}
where \( W[:,:,:,c] \) is the sub-tensor for channel \( c \), and \( C \) is the number of output channels in the weight tensor. The value of $ C $ directly influences the number of unique parameter sets that need to be fetched and applied during inference. This method captures the variability in statistical distributions across channels, enabling more precise quantization by tailoring the scale to each channel's range. For this reason, per-channel quantization generally outperforms per-tensor quantization in architectures such as ResNet, where channel-wise distribution differences are significant.

However, per-channel quantization introduces additional complexity at the hardware level. The additional memory access cost can be approximated as:
\begin{equation}
\text{Cost}_{\text{access}} = C \cdot M_{\text{param}} + T \cdot M_{\text{tensor}}
\label{eq:hardware_cost}
\end{equation}
where \( C \) is the number of channels, \( M_{\text{param}} \) is the memory cost per parameter set. The term $ C \cdot M_{\text{param}} $ estimates the total memory required to store all channel-specific parameter sets, which must be loaded into the hardware's memory (e.g., cache or registers) during inference. \( T \) is the tensor size, and \( M_{\text{tensor}} \) is the memory cost per tensor element. 
The total estimated memory access cost combines cost of loading the parameter sets with the cost of accessing the tensor data. Such memory access can lead to performance bottlenecks if the memory architecture of the hardware accelerator isn't optimized for these non-uniform access patterns.

\subsection{Motivation of Quantized Neural Receiver}
\label{subsec:motivation_ptq}
\begin{figure}[h]
    \centering
    \includegraphics[trim={25 12 5 8},clip, width=0.90\linewidth]
    {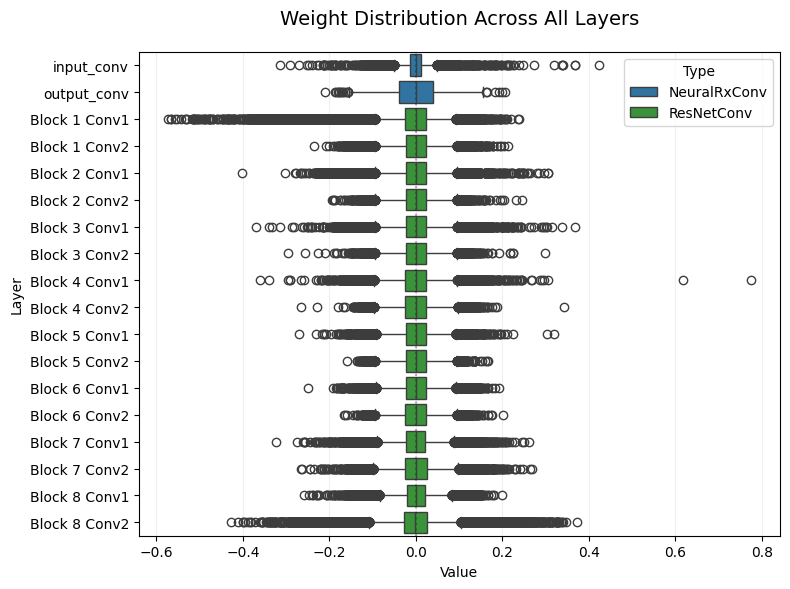}
    \caption{Weight distribution across the layers of the trained neural receiver architecture proposed in this paper. The input and output convolutional layers have a wider spread compared to the compact, zero-centered weight distribution of the ResNetConv layers. Such weight distribution motivates the application of PTQ since the resulting quantization error is expected to be small due to the reduced dynamic range of the ResNetConv layers.}
    \label{fig:weight_distribution}
\end{figure}

The weight distributions across the neural receiver layers reveal a wider spread in input and output convolutional layers compared to the compact, zero-centered distributions of ResNetConv layers. This is illustrated in Fig. \ref{fig:weight_distribution}. Notably, outliers are observed in layers such as Block 4 Conv 1. The ResNetConv layers exhibit a standard deviation of approximately $0.1$, indicating well-regularized weights. This distribution characteristic motivates the application of PTQ, as the reduced dynamic range of ResNetConv layers is expected to minimize quantization error, while the sensitivity of input and output layers necessitates a comprehensive quantization strategy across all layers.\footnote{For this work, input-output layers were quantized uniformly with other convolutional layers.} Experiments in Section \ref{sec:sim_results} will evaluate the impact of varying bit-widths (e.g., 4-bit to 8-bit) on BLER.

\section{Numerical Results}
\label{sec:sim_results}
This section contains numerical results quantifying the performance of the neural receiver architecture described in Section~\ref{subsec:nrx_arch} in terms of BLER (Block Error Rate). The neural receiver architecture was trained using \texttt{float32} precision. Subsequently, PTQ was employed to the weights of the trained model, as discussed in Section~\ref{subsec:motivation_ptq}. The testing parameters are as mentioned in Table \ref{tab:test_parameters}. In particular, the BLER was assessed under three mobility scenarios: low-speed ($0–5$ m/s), medium-speed ($10–20$ m/s), and high-speed ($25–40$ m/s) UE velocities. Moreover, the following receivers have therefore been considered:

\begin{itemize}
    \item \textbf{\texttt{float32} neural receiver}: the full-precision model trained as described in earlier sections.
    
    \item \textbf{Classical receiver}: classical receiver using least squares (LS) channel estimation and nearest-neighbor interpolation, followed by linear minimum mean square error (LMMSE) equalization and soft demapping.
    
    \item \textbf{Ideal receiver with perfect CSI}: classical receiver assuming access to perfect channel state information for equalization and demapping.
    
    \item \textbf{Post-training quantized (PTQ) neural receiver}: the trained \texttt{float32} neural receiver's weights are quantized to lower-precision integer formats, namely \texttt{int8} and \texttt{int4}. Both per-tensor and per-channel quantization schemes are applied to assess their impact on performance under quantized inference.
\end{itemize}

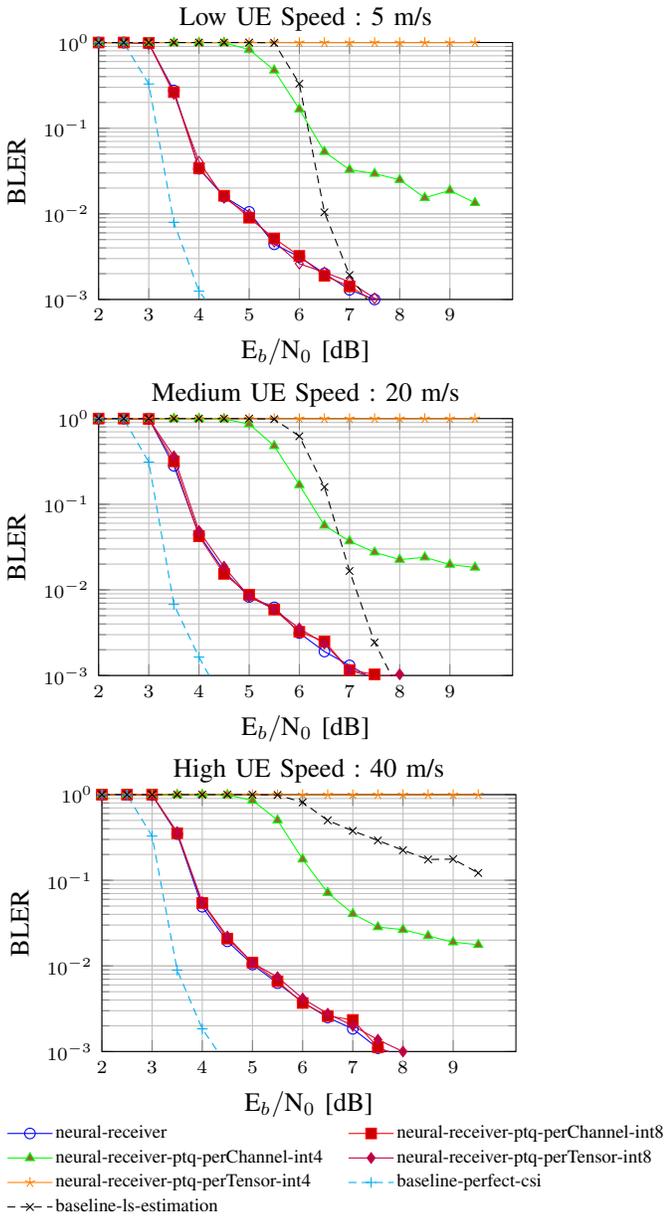
\begin{figure}[h!]
\begin{tikzpicture}
\begin{axis}[
    width=0.8\columnwidth,
    height=5cm,
    xlabel={$\text{E}_b/\text{N}_0$ [dB]},
    ylabel={BLER},
    ymode=log,
    ymin=1e-3, ymax=1, xmin=2,
    grid=both,
    tick label style={font=\scriptsize},
    xtick={2,3,4,5,6,7,8,9},
       title style={yshift=-5pt},
       title = {Low UE Speed : 5 m/s}
]
\addplot+[mark=o, color=blue] coordinates {
(2.0,1.0) (2.5,1.0) (3.0,0.97265625) (3.5,0.27539062) (4.0,0.03359375) (4.5,0.015873017) 
(5.0,0.010526316) (5.5,0.004405286) (6.0,0.00314435) (6.5,0.0020120724) (7.0,0.001296875) 
(7.5,0.001) (8.0,0.000640625) (8.5,0.0005) (9.0,0.00059375) (9.5,0.00053125)
};
\addplot+[mark=square*, color=red] coordinates {
(2.0,1.0) (2.5,1.0) (3.0,0.98828125) (3.5,0.26367188) (4.0,0.034114584) (4.5,0.01624504) 
(5.0,0.009009009) (5.5,0.005154639) (6.0,0.0032258064) (6.5,0.001890625) (7.0,0.001421875) 
(7.5,0.000953125) (8.0,0.000671875) (8.5,0.000734375) (9.0,0.000484375) (9.5,0.000484375)
};
\addplot+[mark=triangle*, color=green] coordinates {
(2.0,1.0) (2.5,1.0) (3.0,1.0) (3.5,1.0) (4.0,1.0) (4.5,0.99609375) 
(5.0,0.828125) (5.5,0.47395834) (6.0,0.16666667) (6.5,0.053042762) (7.0,0.032762095) 
(7.5,0.029464286) (8.0,0.02496189) (8.5,0.015388258) (9.0,0.01875) (9.5,0.0134375)
};
\addplot+[mark=diamond, color=purple] coordinates {
(2.0,1.0) (2.5,1.0) (3.0,0.984375) (3.5,0.25195312) (4.0,0.040625) (4.5,0.015384615) 
(5.0,0.009784588) (5.5,0.0047393367) (6.0,0.002617801) (6.5,0.0020703934) (7.0,0.001609375) 
(7.5,0.00103125) (8.0,0.000921875) (8.5,0.000671875) (9.0,0.000703125) (9.5,0.000546875)
};
\addplot+[mark=star, color=orange] coordinates {
(2.0,1.0) (2.5,1.0) (3.0,1.0) (3.5,1.0) (4.0,1.0) (4.5,1.0) 
(5.0,1.0) (5.5,1.0) (6.0,1.0) (6.5,1.0) (7.0,1.0) (7.5,1.0) 
(8.0,1.0) (8.5,1.0) (9.0,1.0) (9.5,1.0)
};
\addplot+[mark=+, color=cyan] coordinates {
(2.0,1.0) (2.5,0.9921875) (3.0,0.328125) (3.5,0.007935531) (4.0,0.00125) (4.5,0.00053125) 
(5.0,0.000140625) (5.5,4.6875e-05) (6.0,4.6875e-05) (6.5,1e-6) (7.0,1e-6) 
(7.5,1e-6) (8.0,1e-6) (8.5,1e-6) (9.0,1e-6) (9.5,1e-6)
};
\addplot+[mark=x, color=black] coordinates {
(2.0,1.0) (2.5,1.0) (3.0,1.0) (3.5,1.0) (4.0,1.0) (4.5,1.0) 
(5.0,1.0) (5.5,0.9921875) (6.0,0.33072916) (6.5,0.010416667) (7.0,0.0019290124) 
(7.5,0.00075) (8.0,0.0003125) (8.5,0.000125) (9.0,3.125e-05) (9.5,1e-6)
};
\end{axis}
\end{tikzpicture}

\begin{tikzpicture}
\begin{axis}[
    width=0.8\columnwidth,
    height=5cm,
    xlabel={$\text{E}_b/\text{N}_0$ [dB]},
    ylabel={BLER},
    ymode=log,
    ymin=1e-3, ymax=1, xmin=2,
    grid=both,
    tick label style={font=\scriptsize},
    xtick={2,3,4,5,6,7,8,9},
      title style={yshift=-5pt},
    title = {Medium UE Speed : 20 m/s}
]
\addplot+[mark=o, color=blue] coordinates {
(2.0,1.0) (2.5,1.0) (3.0,1.0) (3.5,0.28125) (4.0,0.04415761) (4.5,0.016521517) 
(5.0,0.008260759) (5.5,0.00625) (6.0,0.0031347962) (6.5,0.00190625) (7.0,0.0013125) 
(7.5,0.000890625) (8.0,0.000875) (8.5,0.000453125) (9.0,0.00040625) (9.5,0.000453125)
};
\addplot+[mark=square*, color=red] coordinates {
(2.0,1.0) (2.5,1.0) (3.0,0.98828125) (3.5,0.31640625) (4.0,0.042317707) (4.5,0.015384615) 
(5.0,0.00877193) (5.5,0.00589364) (6.0,0.003236246) (6.5,0.0025) (7.0,0.00115625) 
(7.5,0.00103125) (8.0,0.000796875) (8.5,0.00046875) (9.0,0.00053125) (9.5,0.00025)
};

\addplot+[mark=triangle*, color=green] coordinates {
(2.0,1.0) (2.5,1.0) (3.0,1.0) (3.5,1.0) (4.0,1.0) (4.5,0.98828125) 
(5.0,0.86328125) (5.5,0.47916666) (6.0,0.16796875) (6.5,0.056423612) (7.0,0.037037037) 
(7.5,0.027449325) (8.0,0.022569444) (8.5,0.023995535) (9.0,0.01976103) (9.5,0.018275669)
};

\addplot+[mark=diamond*, color=purple] coordinates {
(2.0,1.0) (2.5,1.0) (3.0,0.98828125) (3.5,0.36197916) (4.0,0.04873512) (4.5,0.018607955) 
(5.0,0.008333334) (5.5,0.00591716) (6.0,0.0035460992) (6.5,0.0023584906) (7.0,0.00115625) 
(7.5,0.0009375) (8.0,0.00103125) (8.5,0.0005) (9.0,0.00053125) (9.5,0.000421875)
};

\addplot+[mark=star, color=orange] coordinates {
(2.0,1.0) (2.5,1.0) (3.0,1.0) (3.5,1.0) (4.0,1.0) (4.5,1.0) 
(5.0,1.0) (5.5,1.0) (6.0,1.0) (6.5,1.0) (7.0,1.0) (7.5,1.0) 
(8.0,1.0) (8.5,1.0) (9.0,1.0) (9.5,1.0)
};

\addplot+[mark=+, color=cyan] coordinates {
(2.0,1.0) (2.5,0.98828125) (3.0,0.31054688) (3.5,0.006756757) (4.0,0.001640625) (4.5,0.000484375) 
(5.0,0.00015625) (5.5,4.6875e-05) (6.0,1e-6) (6.5,1e-6) (7.0,1e-6) 
(7.5,1e-6) (8.0,1e-6) (8.5,1e-6) (9.0,1e-6) (9.5,1e-6)
};

\addplot+[mark=x, color=black] coordinates {
(2.0,1.0) (2.5,1.0) (3.0,1.0) (3.5,1.0) (4.0,1.0) (4.5,1.0) 
(5.0,1.0) (5.5,0.984375) (6.0,0.62109375) (6.5,0.15885417) (7.0,0.016601562) 
(7.5,0.0024338006) (8.0,0.000640625) (8.5,0.000328125) (9.0,9.375e-05) (9.5,9.375e-05)
  };

\end{axis}
\end{tikzpicture}

\begin{tikzpicture}
\begin{axis}[
    width=0.8\columnwidth,
    height=5cm,
    xlabel={$\text{E}_b/\text{N}_0$ [dB]},
    ylabel={BLER},
    ymode=log,
    ymin=1e-3, ymax=1, xmin=2,
    grid=both,
    tick label style={font=\scriptsize},
    title style={yshift=-5pt},
    xtick={2,3,4,5,6,7,8,9},
          legend style={
        font=\scriptsize,
        /tikz/every even column/.append style={column sep=0.3cm},
        draw=none,
        nodes={inner sep=1pt},
        row sep=1pt,
        at={(-0.25,-0.67)},
         anchor=south west,
         legend cell align=left, align=left
    },
    legend columns=2,
    title = {High UE Speed : 40 m/s}
   ]
\addplot+[mark=o, color=blue] coordinates {
(2.0,1.0) (2.5,1.0) (3.0,0.98828125) (3.5,0.34635416) (4.0,0.04910714) (4.5,0.019381009) 
(5.0,0.01038982) (5.5,0.006289308) (6.0,0.0037887688) (6.5,0.0025) (7.0,0.00184375) 
(7.5,0.00109375) (8.0,0.000890625) (8.5,0.00090625) (9.0,0.00090625) (9.5,0.00028125)
};
\addlegendentry{neural-receiver}

\addplot+[mark=square*, color=red] coordinates {
(2.0,1.0) (2.5,1.0) (3.0,1.0) (3.5,0.35416666) (4.0,0.054276317) (4.5,0.020833334) 
(5.0,0.010954483) (5.5,0.0066225166) (6.0,0.003690037) (6.5,0.0025974025) (7.0,0.0023419203) 
(7.5,0.001125) (8.0,0.000828125) (8.5,0.00053125) (9.0,0.000484375) (9.5,0.00040625)
};
\addlegendentry{neural-receiver-ptq-perChannel-int8}

\addplot+[mark=triangle*, color=green] coordinates {
(2.0,1.0) (2.5,1.0) (3.0,1.0) (3.5,1.0) (4.0,1.0) (4.5,0.9921875) 
(5.0,0.85546875) (5.5,0.50390625) (6.0,0.17578125) (6.5,0.071428575) (7.0,0.040625) 
(7.5,0.028428819) (8.0,0.026521381) (8.5,0.022418479) (9.0,0.018952547) (9.5,0.01768092)
};
\addlegendentry{neural-receiver-ptq-perChannel-int4}

\addplot+[mark=diamond*, color=purple] coordinates {
(2.0,1.0) (2.5,1.0) (3.0,0.98828125) (3.5,0.36458334) (4.0,0.055555556) (4.5,0.022078805) 
(5.0,0.010836693) (5.5,0.007410386) (6.0,0.0041645146) (6.5,0.002770083) (7.0,0.002004008) 
(7.5,0.001375) (8.0,0.001) (8.5,0.000875) (9.0,0.00071875) (9.5,0.000640625)
};
\addlegendentry{neural-receiver-ptq-perTensor-int8}

\addplot+[mark=star, color=orange] coordinates {
(2.0,1.0) (2.5,1.0) (3.0,1.0) (3.5,1.0) (4.0,1.0) (4.5,1.0) 
(5.0,1.0) (5.5,1.0) (6.0,1.0) (6.5,1.0) (7.0,1.0) (7.5,1.0) 
(8.0,1.0) (8.5,1.0) (9.0,1.0) (9.5,1.0)
};
\addlegendentry{neural-receiver-ptq-perTensor-int4}

\addplot+[mark=+, color=cyan] coordinates {
(2.0,1.0) (2.5,0.98828125) (3.0,0.328125) (3.5,0.008928572) (4.0,0.00184375) (4.5,0.0006875) 
(5.0,0.000171875) (5.5,0.000109375) (6.0,1e-6) (6.5,1e-6) (7.0,1e-6) 
(7.5,1e-6) (8.0,1e-6) (8.5,1e-6) (9.0,1e-6) (9.5,1e-6)
};
\addlegendentry{baseline-perfect-csi}

\addplot+[mark=x, color=black] coordinates {
(2.0,1.0) (2.5,1.0) (3.0,1.0) (3.5,1.0) (4.0,1.0) (4.5,1.0) 
(5.0,1.0) (5.5,0.9921875) (6.0,0.8125) (6.5,0.5) (7.0,0.37760416) 
(7.5,0.29166666) (8.0,0.22460938) (8.5,0.175) (9.0,0.1765625) (9.5,0.12165178)
};
 \addlegendentry{baseline-ls-estimation}
\end{axis}
\end{tikzpicture}
\vspace{-0.3cm}
\caption{Performance of the proposed neural receiver architecture in terms of BLER versus energy-per-bit to noise-power-spectral-density ratio $E_b/N_0$. The performance of the classical receiver is given as well. With practical LS channel estimation and with ideal CSI. Both are used as baseline. Performance in low UE speed scenarios is given top panel. Medium and high UE speed scenarios are given in middle and botton panels, respectively. The proposed neural receiver architecture with PTQ using \texttt{int8} precision outperforms the classical receiver with practical LS channel estimation in all mobility scenarios. Improvements range from $2$dB to $5$dB at $10\%$ BLER.\label{fig:bler_plots}}
\end{figure}

Figure \ref{fig:bler_plots} illustrates the performance of the proposed neural receiver architecture in terms of BLER versus energy-per-bit to noise-power-spectral-density ratio $E_b/N_0$. The performance of the classical receiver is given, both with practical LS channel estimation as well as with ideal CSI. Both are used as baseline. Assessed scenarios include low UE speed in (a), medium UE speed in (b), and high UE speed in (c). The proposed neural receiver architecture with PTQ using \texttt{int8} precision outperforms the classical receiver with practical LS channel estimation in all mobility scenarios. Improvements range from $2$dB to $5$dB at $10\%$ BLER. Concretely, the proposed neural receiver architecture using \texttt{int8} quantization achieves a performance comparable with using \texttt{float32} precision. Both per-tensor and per-channel PTQ using \texttt{int8} precision exhibit similar performance.

Per-channel PTQ with 4-bit integer precision outperforms the classical receiver with practical LS channel estimation at $10\%$ BLER for increasing UE speed. Per-tensor PTQ with 4-bit integer precision leads to a significant performance degradation of the neural receiver architecture across all scenarios considered in this numerical study. The models size reduction after performing PTQ is summarized in \ref{tab:quant-size-summary}.
 
\begin{table}[htbp]
\centering
\begin{threeparttable}
\caption{Neural Receiver Variants: Bit-width and Model Size Reduction}
\label{tab:quant-size-summary}
\begin{tabular}{|l|c|c|}
\hline
\textbf{Receiver Variant} & \textbf{Bit-width} & \textbf{Model Size Reduction} \\
\hline
\texttt{float32} Neural Receiver & 32-bit & Baseline (1$\times$) \\
\hline
PTQ Per-Channel \texttt{float16} & 16-bit & $\approx$2$\times$ \\
PTQ Per-Channel \texttt{int} & 8-bit & $\approx$4$\times$ \\
PTQ Per-Channel \texttt{int} & 4-bit & $\approx$8$\times$ \\
PTQ Per-Tensor \texttt{int} & 8-bit & $\approx$4$\times$ \\
PTQ Per-Tensor \texttt{int} & 4-bit & $\approx$8$\times$ \\
\hline
\end{tabular}
\begin{tablenotes}
\footnotesize 
\item \textbf{Note}: Model size reduction is approximate, based on weight quantization only, as LayerNormalization parameters remain in \texttt{float}.
\end{tablenotes}
\end{threeparttable}
\end{table}

\section{Conclusion and Future Work} \label{sec:conclusions}
A deep neural network receiver architecture has been proposed. Training was carried out using \texttt{float32} precision. Post-training quantization (PTQ) of the model weights using \texttt{int8} and \texttt{int4} precision have been considered. Link-level numerical results using Sionna have shown that the proposed receiver architecture with PTQ using \texttt{int8} precision achieves similar performance to that of \texttt{float32} precision. This suggests that \texttt{int8} PTQ can be considered for neural network wireless receivers for real-world deployment. The proposed architecture with PTQ using \texttt{int8} precision outperforms the classical (non-DL) receiver in terms of energy-per-bit to noise-power-spectral-density ratio at $10\%$-BLER by $2-5$ dB on various mobility scenarios.

The proposed architecture with per-channel PTQ using \texttt{int4} precision has been shown to outperform the classical (non-DL) receiver in high mobility scenarios. Future work will focus on \texttt{int4} precision PTQ performance on different training strategies under varying mobility conditions. Another important direction for us is to experiment with quantization-aware training, both from scratch and with a warm start. This is essential for enhancing the feasibility of deploying deep receiver architectures in resource-constrained environments, ultimately improving energy efficiency and enabling real-time inference.

\section*{Acknowledgements}
The authors thank Sebastian Cammerer from NVIDIA for his insights during the inception of this work. This work has been supported in parts by Research Council of Finland (grant no:359848) and by the 6GARROW project which has received funding from the Smart Networks and Services Joint Undertaking (SNS JU) under the European Union’s Horizon Europe research and innovation programme under Grant Agreement No 101192194 and from the Institute for Information \& Communications Technology Promotion (IITP) grant funded by the Korean government (MSIT) (No. RS-2024-00435652).

\end{document}